\documentclass[conference]{IEEEtran}
\IEEEoverridecommandlockouts
\usepackage{cite}
\usepackage{amsmath,amssymb,amsfonts}
\usepackage{algorithmic}
\usepackage{graphicx}
\usepackage{textcomp}
\usepackage{url}
\usepackage{xcolor}
\def\BibTeX{{\rm B\kern-.05em{\sc i\kern-.025em b}\kern-.08em
    T\kern-.1667em\lower.7ex\hbox{E}\kern-.125emX}}

\usepackage{multirow}
\usepackage{makecell}

\begin{document}
\title{Autoencoder based optimized SSL representations: Complexity Minimization and improved Dysarthric ASR}


\author{
    \IEEEauthorblockN{Paban Sapkota\IEEEauthorrefmark{1}, Hemant Kumar Kathania\IEEEauthorrefmark{1},
    Mikko Kurimo\IEEEauthorrefmark{2},
    Shrikanth Narayanan\IEEEauthorrefmark{3}, and Sudarsana Reddy Kadiri\IEEEauthorrefmark{3}}
    \\
    \IEEEauthorblockA{\IEEEauthorrefmark{1} Department of Electronics and Communication Engineering, National Institute of Technology Sikkim, India.  \\
    Emails: phec230006@nitsikkim.ac.in, hemant.ece@nitsikkim.ac.in}

    \IEEEauthorblockA{\IEEEauthorrefmark{2} Department of Information and Communications Engineering, Aalto University, Finland.\\
Email: mikko.kurimo@aalto.fi}
    \IEEEauthorblockA{\IEEEauthorrefmark{3} Signal Analysis and Interpretation Laboratory (SAIL), University of Southern California, Los Angeles, USA.\\
    Emails: shri@usc.edu, skadiri@usc.edu}
}

\maketitle

\begin{abstract}

Self-supervised learning (SSL) models extract rich speech representations but often come with high-dimensional features, increasing computational complexity. This work explores an SSL-AutoEncoder (SSL-AE) bottlenecking approach to efficiently reduce feature dimensions while maintaining dysarthric Automatic Speech Recognition (ASR) performance. By leveraging an autoencoder, we transform high-dimensional SSL features into a compact space, reducing model complexity and training time. Our method preserves essential speech information, achieving reduced Word Error Rates (WER) while significantly lowering computational costs. Experiments show SSL-AE bottlenecking reduces training time by 8× compared to the SSL baseline, demonstrating efficiency without sacrificing recognition performance. These results highlight AE as an effective solution for SSL feature compression in resource-constrained environments.

\end{abstract}

\vspace{-0.3cm}
\section{Introduction}\vspace{-0.1cm}
\label{sec:intro}
Speech recognition technologies have seen remarkable advancements over the years. The transition from conventional Hidden Markov Models (HMMs) to more modern end-to-end frameworks, coupled with the rise of large, pre-trained self-supervised learning (SSL) models, has greatly enhanced the performance of Automatic Speech Recognition (ASR) systems. These improvements have similarly impacted dysarthric speech recognition, leading to the development of more precise and resilient models for interpreting speech of individuals affected by dysarthria.

Dysarthric speech is typically marked by slow, unclear, and variable articulation. Many ASR systems have been proposed in the literature to address these challenges. While conventional systems, such as those relying on Hidden Markov Models (HMM) and Gaussian Mixture Models (GMM), have been commonly applied, they face difficulties due to the unpredictable and inconsistent nature of dysarthric speech \cite{qian2023survey}. To mitigate these issues, hybrid approaches combining HMM-GMM with Deep Neural Networks (DNNs) have been explored, resulting in enhanced recognition accuracy.

In recent years, end-to-end speech recognition models leveraging architectures like Convolutional Neural Networks (CNNs), Recurrent Neural Networks (RNNs) \cite{almadhor2023e2e,bhat2022improved,takashima2019end,yue2022acoustic}, and Transformer-based networks \cite{ding2021multi,sinha2024effect,shahamiri2023dysarthric} have gained traction. These models excel at capturing the complexities of dysarthric speech by learning robust feature representations directly from data \cite{wang2024enhancing}. Additionally, transfer learning with pre-trained models such as Wav2Vec2 \cite{baveski}, HuBERT \cite{hubert}, and Data2Vec \cite{data2vec} has been explored, demonstrating significant improvements when fine-tuned on dysarthric speech. By leveraging knowledge from large-scale general speech corpora, these models enhance recognition accuracy \cite{wang2024enhancing}. Furthermore, customized ASR approaches, incorporating speaker-adaptive models and advanced language modeling techniques, have been developed to better accommodate the unique characteristics of dysarthric speech.

In this paper, we explore the utilization of features extracted from large pretrained SSL models in traditional hybrid DNN-HMM systems. Our investigation examines whether the SSL model embeddings are suitable for use in a Kaldi-based \cite{kaldi,hermann,xiong1} ASR pipelines \cite{10983605}. We also propose an AutoEncoder (AE) \cite{yue2020autoencoder, vachhani2017deep} block for SSL feature dimension optimization, integrating pre-trained model features into the ASR pipeline in an efficient manner.
Following are the key contributions of this paper:\begin{itemize}
    
    \item Developed a Kaldi-based DNN-HMM ASR pipeline for dysarthric speech using pretrained SSL model embeddings, achieving improved efficiency compared to the MFCC baseline system.

    \item Evaluated direct decoding and fine-tuning of pretrained SSL models, investigating severity-independent and speaker-independent ASR performance for dysarthric speech.

    \item Proposed an autoencoder-based SSL representation optimization method using HuBERT embeddings to reduce feature dimensionality and improve ASR performance. 
\end{itemize}

\section{Automatic Speech Recognition (ASR) Pipeline}
\label{sec:expsetup}
\begin{figure*}[ht]
  \centering
  \includegraphics[width=\linewidth]{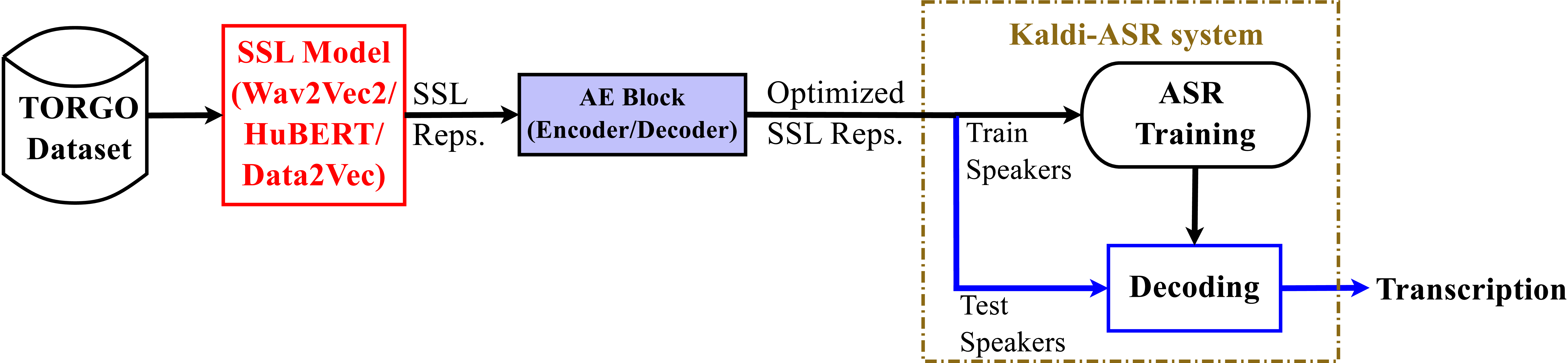}
  \caption{Block diagram of the proposed kaldi ASR pipeline for dysarthric speech recognition. `Reps.' abbreviates for representations.}
  \label{fig:speech_production}
\end{figure*}

We utilize SSL embeddings as input features in a Kaldi-based DNN-HMM ASR system. Between the feature extraction and the ASR model, we introduce an AutoEncoder (AE)-based feature dimension optimization block. An overview of the ASR pipeline is provided in Figure \ref{fig:speech_production}.

\subsection{Dataset and Training/Testing Protocol}
Only a few dysarthric speech corpora are publicly available. Most existing datasets consist of single-word utterances, limiting their effectiveness for ASR training. The TORGO dataset  \cite{torgo} is one of the few publicly available corpora that include sentenced utterances, which are crucial for building robust ASR systems. Sentenced utterances help ASR models learn contextual dependencies, prosody, and natural speech patterns, improving recognition performance.The TORGO corpus comprises roughly 15 hours of speech recordings collected from 15 speakers. This set includes 8 speakers with dysarthria and 7 age-matched control speakers without speech impairments. Speech data from all 15 speakers are used in the present study. The preprocessing procedure follows the methodology outlined by Hermann et al. \cite{hermann}, which involved eliminating empty audio files, those with complex transcriptions, and files lacking transcriptions. Additionally, utterances with fewer than 35 frames were excluded from the dataset. The technical details about the dataset can be found in  \cite{torgo}.

Dysarthria severity levels for the speakers in the dataset were assigned according to criteria established in prior studies \cite{9762324,sapkota2024improving}. Depending on the severity of the speaker, we divided the speakers into four groups. One control group (Ctl.) consisted of the seven control speakers. The eight dysarthric speakers were assigned to three severity groups: three speakers were placed in the low severity (Low) group, two in the medium severity (Med) group, and three in the high severity (High) group. The corresponding speaker-IDs and the number of utterances used in this study are tabulated for clarity in Table \ref{tab:speaker_stats}.

\renewcommand{\arraystretch}{1.2}
\begin{table}[h]
    \centering
    \caption{Speaker details in the TORGO dataset with severity labels, followed by data distribution for the train-test protocols. `Tr' and `TE' represents the data groups being used while training and testing, respectively.} \vspace{-0.2cm}
    \label{tab:speaker_stats}
    \resizebox{0.48\textwidth}{!}{%
    \begin{tabular}{ccccc}\hline
         \multicolumn{5}{c}{\bf{Severity labels and utterance statistics}}\\\hline
        Severity: & Low & Med & High & Ctl. \\
         \multirow{2}{*}{Spk-ID:} &  \multirow{2}{*}{F03, F04, M03} &  \multirow{2}{*}{F01, M05} &  \multirow{2}{*}{M01, M02, M04}  &  \multirow{2}{*}{\shortstack{FC01, FC02, FC03, \\  MC01, MC02, MC03, MC04}}  \\
         &&&& \\
        \#Utts.: &2532 & 801 & 2156 & 10892 \\\hline
        \multicolumn{5}{c}{\bf{Training/Testing system settings}}\\\hline
        Sys0:&TE&TE&TE& Tr\\
        Sys1:&Tr&TE&TE& Tr\\
        Sys2:&TE&Tr&TE& Tr\\
        Sys3:&TE&TE&Tr& Tr\\ \hline
    \end{tabular}
}    
\end{table}   

Since the dataset lacks a predefined train-test split, we partitioned the data by dysarthric severity label for balanced evaluation. Most experiments hereafter follow four train-test systems (Sys0, Sys1, Sys2, and Sys3) as shown in Table \ref{tab:speaker_stats}, implementing both severity-independent and speaker-independent strategies.

\subsection{Description of the ASR System}
The ASR system utilizes DNN-HMM hybrid system, implemented using the Kaldi toolkit \cite{kaldi}, for acoustic modelling. A 2-gram language model is used throughout the study, for decoding and constraining word sequences during recognition, using SRILM toolkit \cite{stolcke2002srilm}. The training process began with monophone models, which were later refined into triphone models. These models incorporated LDA-MLLT (Linear Discriminant Analysis and Maximum Likelihood Linear Transform) and SAT (Speaker Adaptive Training) to account for speaker-specific variations \cite{sat}. Speaker-dependent transforms are first applied to obtain alignments from the HMM–GMM system, after which the final acoustic model is trained using a DNN–HMM hybrid framework. The DNN comprises three hidden layers, each with 300 neurons. It's training employs a learning rate schedule that starts at 0.04 and is gradually decreased to 0.004. Training is carried out for 15 epochs, with a further 5 epochs for fine-tuning. A mini-batch size of 128 and 16 parallel jobs were used to speed up training. Detailed layer configurations and hyperparameter settings can be found in \cite{9053334,10942704}.

\subsection{SSL Model Details}
\label{sec:SSL_detail}
Three comparable SSL speech models were selected for this study, all sharing the same architectural design with a seven-layer convolutional feature encoder and a twenty-four-layer transformer-based context encoder. These models include Wav2Vec2 \cite{baveski}, HuBERT \cite{hubert}, and Data2Vec \cite{data2vec}, proposed by Meta AI (formerly FAIR).
The specific models used are wav2vec2-large-960h-lv60-self, hubert-large-ls960-ft, and data2vec-audio-large, availabe in huggingface repository. The embeddings from the final transformer layers are extracted, as they serve as the basis for decoding in the respective SSL models. The feature extraction process from the final transformer layer of a SSL model involves passing raw speech waveforms through the pretrained model and retrieving embeddings from the last hidden layer, 1024 embeddings per 20ms of speech frame. The extracted features (SSL reps.) are then stored in a Kaldi-compatible format for further processing, using kaldiio python library.
\section{Proposed AutoEncoder (AE) network for optimizing SSL reps.}
\label{sec:AE}
\begin{figure}[h]
  \centering
  \includegraphics[width=\linewidth]{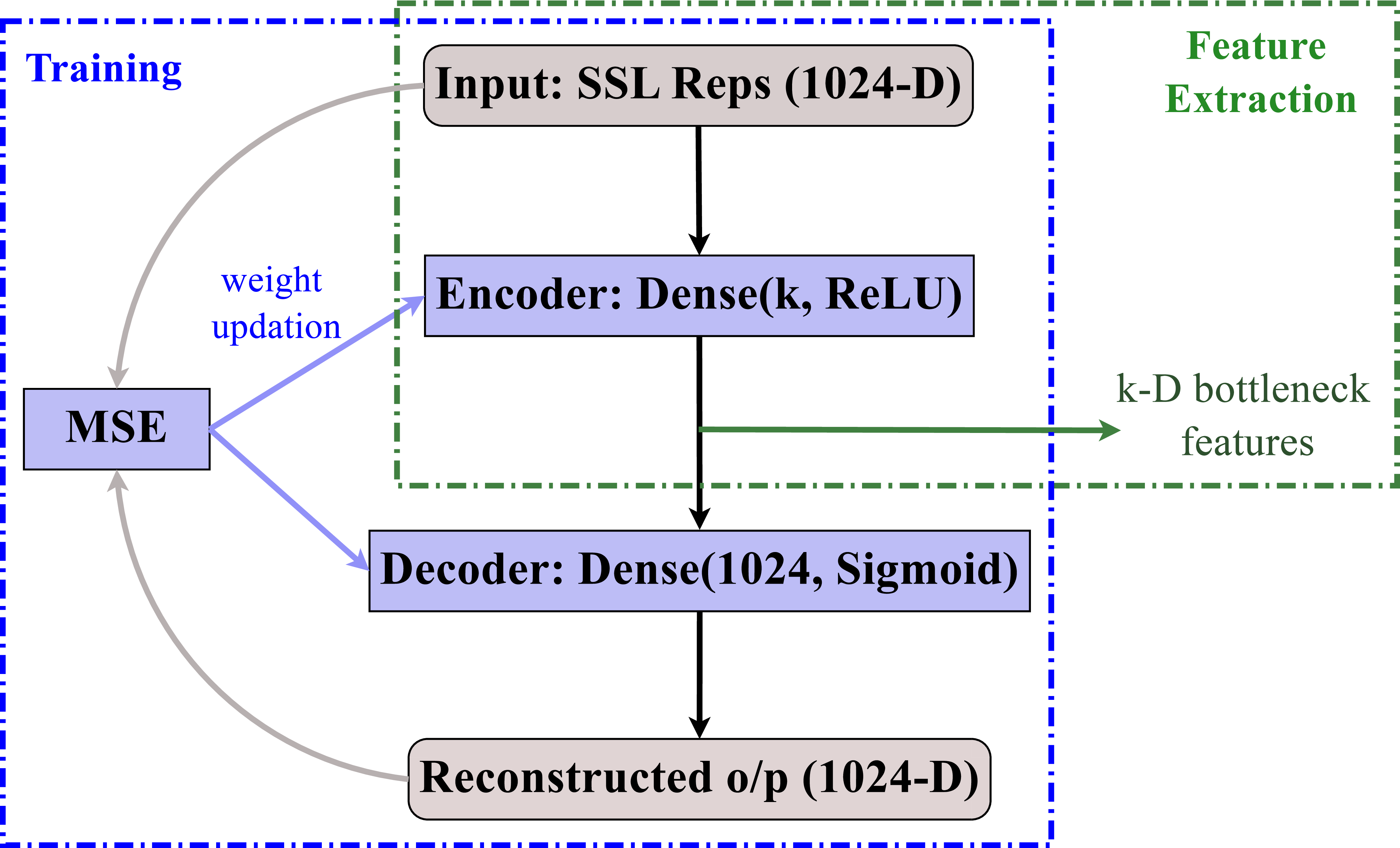}
  \caption{Block diagram of the proposed autoencoder (AE) architecture. The `Training' module illustrates the complete network employed during model optimization, while the `Feature Extraction' module highlights the subnetwork used to obtain bottleneck representations. Here, `Reps' denotes feature representations, and k represent the dimensions of the bottleneck encoding.}
  \label{fig:AE_system}
\end{figure}

The autoencoding process follows an unsupervised learning approach using a simple Encoder-Decoder neural network. The network is trained using the Adam optimizer for 20 epochs per batch, aiming to minimize the reconstruction loss between the input features and the decoder’s output. The training is performed batch-wise, where 256 frames of 1024-dimensional input features are processed at a time. Each batch undergoes 20 epochs before moving to the next batch, ensuring incremental learning across the dataset. 
A visual representation of the autoencoder architecture is provided in Figure \ref{fig:AE_system}.

After the training is complete, only the encoder part of the network is used for extracting bottleneck features, eliminating the need for the decoder. The encoder transforms the high-dimensional input features into a lower-dimensional representation, with k = 512, 256, 128, 64, 32, and 13 being experimented with. Finally, the extracted bottleneck features are stored in kaldi-readable formats (.ark and .scp), ensuring compatibility with subsequent ASR pipelines.

\section{Results}

 Three baseline systems were developed: (1) a Kaldi-ASR system with 13D MFCCs, (2) direct inference using SSL models (as mentioned in Section \ref{sec:SSL_detail}), and (3) SSL fine-tuning baselines (Section \ref{sec:ft-base}). Section \ref{sec:SSL_embed} presents results using SSL embeddings in DNN-HMM ASR, while Section \ref{sec:AE_optimized} details the proposed AE-optimized SSL representations.
 
\subsection{MFCC and SSL decoding baselines}
\label{sec:mfcc_SSLdecode}
The evaluation of the baseline ASR systems were conducted for the four training–testing configurations described in Table~\ref{tab:speaker_stats} (Sys0–Sys3). These results include decoding using 13-dimensional MFCC features with the models trained using Kaldi ASR (DNN-HMM) framework, as well as direct zero-shot inference utilizing the pretrained SSL models. The corresponding results are summarized in Table~\ref{tab:merged_results}.
\renewcommand{\arraystretch}{1.2}
\begin{table}[h]
\centering
\caption{WER (\%) across severity levels. Upper panel: MFCC-based DNN-HMM baselines under different training configurations. Lower panel: zero-shot ASR performance of SSL models without fine-tuning. ``Tr'' denotes severity levels included during training.}
\vspace{-0.2cm}
\label{tab:merged_results}
\resizebox{0.44\textwidth}{!}{%
\begin{tabular}{lccccc}
\hline
\multicolumn{6}{c}{\textbf{MFCC-based DNN--HMM Baselines}} \\ \hline
System & Training & Low & Medium & High & Avg. \\ \hline
Sys0 & MFCC & 23.20 & 115.41 & 104.86 & \multirow{4}{*}{55.89} \\
Sys1 & MFCC & Tr & 95.89 & 86.77 &  \\
Sys2 & MFCC & 22.93 & Tr & \textbf{80.06} &  \\
Sys3 & MFCC & \textbf{21.93} & \textbf{65.69} & Tr &  \\ \hline
\multicolumn{6}{c}{\textbf{SSL Models (Zero-shot Decoding)}} \\ \hline
Model & -- & Low & Medium & High & Avg. \\ \hline
Wav2Vec2 & -- & 30.15 & 68.03 & 82.87 & 60.35 \\
HuBERT   & -- & 26.83 & 72.88 & 91.94 & 63.88 \\
Data2Vec & -- & 30.09 & 71.26 & 89.55 & 63.63 \\ \hline
\end{tabular}
}
\end{table}

The MFCC results indicate that training solely with control data (Sys0) leads to poor WERs for Medium severity as well as the High severity group of speakers. Recognition improves with Sys2 and Sys3, where training includes Medium or High severity data. The ‘Avg.’ represents the best system's average WER across severity groups (Avg. WER: 55.89\%). In SSL model inference, Wav2Vec2 outperformed HuBERT and Data2Vec, achieving the best average WER (60.35\%) and performing superiorly in the Medium and High severity test groups. Although these experiments are not directly comparable, the results suggest that supervised training with dysarthric speech is beneficial, as large pretrained models may have limited exposure to such speech. This observation prompts an investigation into fine-tuning the pretrained model for a more balanced comparison, as presented in Section \ref{sec:ft-base}.

\subsection{SSL fine-tuning baseline}
\label{sec:ft-base}
Fine-tuning was conducted solely on the Wav2Vec2 model, selected based on direct inference results (Table \ref{tab:merged_results}), to assess ASR performance on dysarthric speech. Three sets of fine-tuning experiments were conducted, each using training data from a specific severity level alongside control speech data, as shown in Table \ref{tab:TORGO_severity_wise}. 
\begin{table}[h]
    \centering
    \caption{Severity-Specific and Speaker Independent performance evaluation of TORGO dysarthric speech in terms of WER (\%). Severity levels `Low', `Med', and `High' denote speech from the specific severity categories within the TORGO corpus.}
    \vspace{-0.2cm}
    \label{tab:TORGO_severity_wise}
\resizebox{0.3\textwidth}{!}{ 
    \begin{tabular}{cccc} \hline 
         \bf{Training data}&  \multicolumn{3}{c}{\bf{Test Severity}}\\\cline{2-4} \bf{for Fine-tuning}
 & Low& Medium& High\\\hline \hline 
         Sys1&  Tr&  80.13&  79.38\\ 
         Sys2&  14.94&  Tr&  \bf{65.24}\\  
         Sys3&  \bf{13.92}&  \bf{43.28}&  Tr\\ \hline
         Avg. & \multicolumn{3}{c}{40.48} \\\hline
    \end{tabular}
}       
\end{table} 

We can observe that the fine-tuning results are significantly better than the MFCC baseline on kaldi ASR. The Avg. of the best systems for the severity in test averaged to 40.48\%, which is 15.41\% absolutely better than the DNN-HMM baseline 1 (in Table \ref{tab:merged_results}). This comes at the cost of high GPU usage (11 Gibibytes (GiB)) and approximately 30 hours of training on a dedicated A5000 16 GiB desktop GPU running Ubuntu. Thus, a better approach could combine the strengths of both worlds, supervised ASR in Kaldi and SSL model embeddings as features.


\subsection{SSL embeddings as features}
\label{sec:SSL_embed}

The final transformer layer embeddings (1024D) from the three SSL models replaced the 13D MFCC features in the DNN-HMM ASR system. Table \ref{tab:dnn_T} presents the results for the four training-testing systems (Sys0 to Sys3). 

\renewcommand{\arraystretch}{1.2}
\begin{table}[h]\vspace{-0.5cm}
    \centering
    \caption{Comparison of DNN results across different train-test systems and SSL models for feature extraction. The table presents Word Error Rates (WERs) for different test severities: Low, Medium (Med), and High. 'Tr' indicates unavailable test results.} \vspace{-0.2cm}
    \label{tab:dnn_T}
    \resizebox{0.4\textwidth}{!}
    {%
        \begin{tabular}{ccccc}
        \hline
             \multirow{2}{*}{\bf{Train Set}}&\multirow{2}{*}{\bf{Feature}}&\multicolumn{3}{c}{\bf{Test Severity}}  \\\cline{3-5}
             &&Low&Med&High \\ \hline
             
             \multirow{3}{*}{Sys0}&Wav2Vec2&14.55&44.20&57.31\\
             &HuBERT&11.55&40.19&54.34\\
             &Data2Vec&13.99&43.28&57.53\\\hline
             
             \multirow{3}{*}{Sys1}&Wav2Vec2&Tr&43.84&52.73\\
             &HuBERT&Tr&36.70&48.68\\
             &Data2Vec&Tr&38.11&50.26\\\hline
             
             \multirow{3}{*}{Sys2}&Wav2Vec2&12.87&Tr&48.40\\
             &HuBERT&10.95&Tr&\bf{42.18}\\
             &Data2Vec&13.63&Tr&44.53\\\hline

             \multirow{3}{*}{Sys3}&Wav2Vec2&12.95&30.66&Tr\\
             &HuBERT&\bf{10.84}&\bf{27.83}&Tr\\
             &Data2Vec&13.66&29.09&Tr\\
             \hline
             Avg. &&\multicolumn{3}{c}{26.95}\\\hline
             
        \end{tabular}
    }
\end{table}

All four systems (Sys0 to Sys3) showed significantly better performance with HuBERT embeddings compared to Wav2Vec2 and Data2Vec. Among the evaluated models, HuBERT consistently produced the strongest recognition results across all severity categories, yielding WERs of 10.84\%, 27.83\%, and 42.18\% for the Low, Medium, and High test sets, respectively. This corresponds to an overall average WER of 26.95\%, representing a relative improvement of 13.47\% compared to the fine-tuned baseline. HuBERT outperformed other SSL representations, likely due to its unique pretraining strategy, which leverages clustering-based feature learning for improved phonetic modeling. The main drawback of using SSL embeddings as features is the high computational cost due to their large feature dimensions.

\subsection{AE optimized SSL features}
\label{sec:AE_optimized}

\begin{figure}[h]\vspace{-0.2cm}
  \centering
  \includegraphics[width=\linewidth]{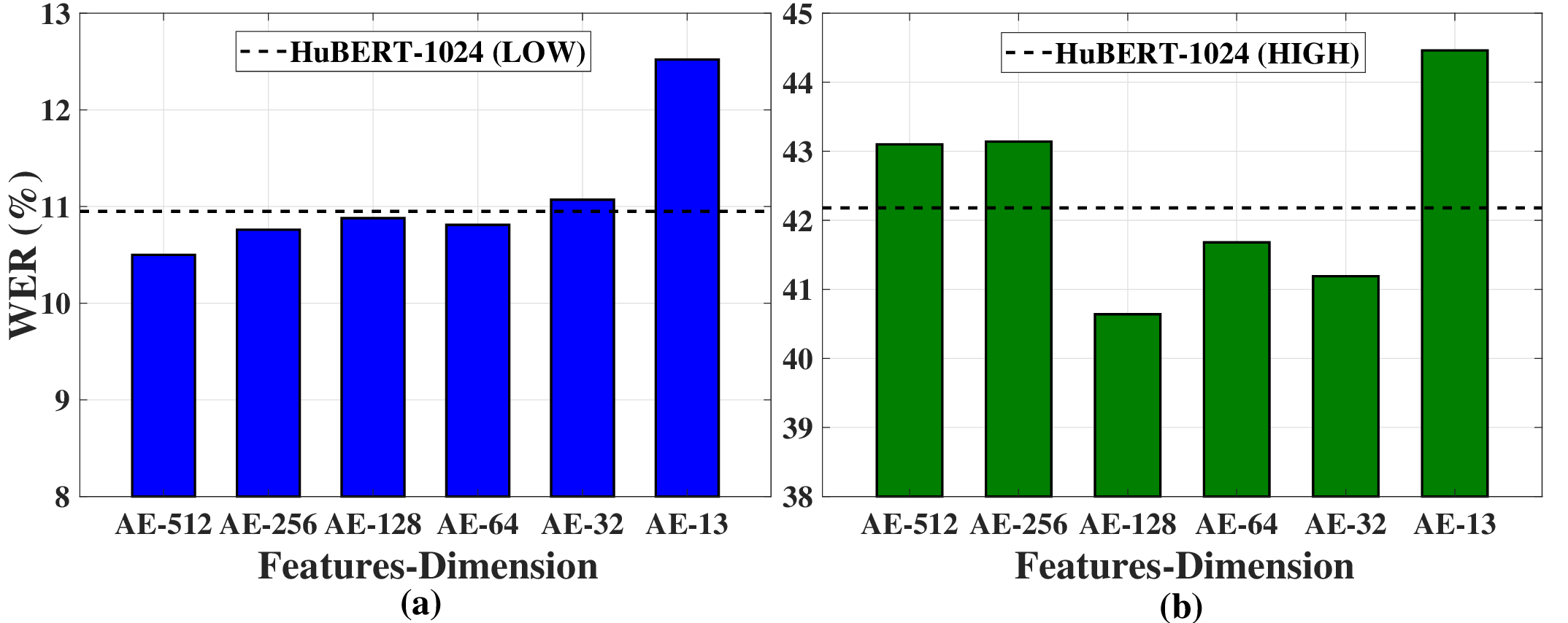}
  \caption{Bar plot showing Word Error Rate (WER) for (a) Low severity and (b) High severity evaluations under varying feature dimensional settings. The visualization compares WER across AE-optimized feature dimensions under the two severity conditions.}
  \label{fig:plot_AE_variations}
\end{figure}

To mitigate the issue caused by the larger dimension of embeddings from the SSL model, we adopt the AE-based feature optimization strategy described in Section~\ref{sec:AE}. Experiments on Sys2, which is suitable for both Low- and High-severity test conditions, while progressively reducing the feature dimension from k=512 to k=13, as depicted in Fig. \ref{fig:plot_AE_variations}.
\vspace{-0.1cm}

For high-severity test speech, Fig.~\ref{fig:plot_AE_variations} suggests that the AE-128 bottleneck representation yields lowest WER in comparison to the original 1024-dimensional HuBERT features, whereas the lowest WER for Low severity is obtained at k=512. Since the performance gain is more substantial in the High-severity condition, AE-128 is selected for comprehensive evaluation on Sys0 through Sys3, as summarized in Table \ref{tab:hubert_ae_comparison}.
\vspace{-0.4cm}
\renewcommand{\arraystretch}{1.15}
\begin{table}[h]
\centering
\caption{Severity-wise WERs (\%) for different training configurations using full-dimensional HuBERT features and autoencoder-compressed representations. Boldface indicates the lowest WERs.}
\label{tab:hubert_ae_comparison}
\resizebox{0.48\textwidth}{!}{%
\begin{tabular}{ccccc} \hline
\bf{Model} & \bf{Train Set} & \bf{Low} & \bf{Medium} & \bf{High} \\ \hline
HuBERT-1024 & Sys0 & 11.55 & 40.19$^{\dagger}$ & 54.34 \\
AE-128      & Sys0 & 11.36$^{\dagger}$ & 41.36 & 53.71$^{\dagger}$ \\ \hline
HuBERT-1024 & Sys1 & Tr & 36.70 & 48.68 \\
AE-128      & Sys1 & Tr & 33.76$^{\dagger}$ & 42.23$^{\dagger}$ \\ \hline
HuBERT-1024 & Sys2 & 10.95 & Tr & \bf{42.18} \\
AE-128      & Sys2 & 10.88$^{\dagger}$ & Tr & \bf{40.62}$^{\dagger}$ \\ \hline
HuBERT-1024 & Sys3 & \bf{10.84} & \bf{27.83} & Tr \\
AE-128      & Sys3 & \bf{10.28}$^{\dagger}$ & \bf{27.01}$^{\dagger}$ & Tr \\ \hline
\multicolumn{2}{c}{Average} & \multicolumn{3}{c}{26.95 (HuBERT) \quad / \quad 25.97 (AE)} \\ \hline
\vspace{-0.15cm}
\end{tabular}
}

\footnotesize{`$\dagger$' indicates the lower WER between HuBERT-1024 and AE-128 models.}
\end{table}

Table \ref{tab:hubert_ae_comparison} reports a comparison of HuBERT-1024 and optimized AE-128 features, indicating that AE-128 achieves superior performance across most of the train–test configurations (Sys0–Sys3). In particular, the best-performing system under High-severity conditions attains an absolute reduction in WER by 1.56\%, primarily due to fewer deletion and substitution errors. Averaged across systems, AE-128 achieves a WER of 25.97\%, corresponding to a 0.98\% absolute improvement over HuBERT-1024.

These gains may be partly attributable to capacity limitations in one or both of the AE and ASR architectures, which warrant further analysis. Reduced feature dimensionality can improve ASR generalization and robustness by suppressing noise, mitigating overfitting, and stabilizing acoustic–phonetic alignment, thereby lowering WER. Conversely, high-dimensional representations often include redundant information, leading to increased computational cost, higher overfitting risk, and diminished overall performance.
 
\section{Discussion}

The experimental findings indicate that, for dysarthric speech recognition, integrating SSL representations within a conventional DNN-HMM based ASR framework yields more reliable performance than both zero-shot decoding with pretrained SSL models and direct fine-tuning of the SSL front end. This suggests that SSL embeddings are more effectively exploited when used as feature representations rather than as end-to-end decoders in low-resource and atypical speech conditions.

From a computational perspective, this hybrid approach also offers clear advantages. In particular, compressing high-dimensional HuBERT representations using an AE leads to a substantial reduction in model complexity while preserving recognition accuracy. As a result, the AE-based feature optimization proves to be a more practical choice for large-scale or resource-constrained ASR training.

The computational cost associated with training and inference is reported in Table~\ref{tab:hubert_ae_time_comparison}. The results show a pronounced decrease in training time when AE-128 bottleneck features are employed instead of the original HuBERT-1024 representations.
 \vspace{-0.2cm}

\renewcommand{\arraystretch}{1.15}
\begin{table}[h]
\centering
\caption{System duration comparison of HuBERT-1024 and AE-128 during ASR training and testing.}
\label{tab:hubert_ae_time_comparison}
\resizebox{0.3\textwidth}{!}{%
\begin{tabular}{cccc} \hline
\bf{Feature} & \bf{Phase} & \bf{Time (min)} \\ \hline
\multirow{2}{*}{HuBERT-1024} & Training & 199 \\
 & Testing  & 11.33 \\ \hline
\multirow{2}{*}{AE-128}      & Training & 25 \\
 & Testing  & 10.38 \\ \hline
\end{tabular}
}
\end{table}

Specifically, the use of AE-128 features reduces ASR training time by nearly an 8x order of magnitude compared to the full-dimensional HuBERT features. Although the AE itself requires additional training and feature extraction time, this overhead amounts to only 14 minutes. When combined with the ASR training time, the total duration is approximately 39 minutes, which remains substantially lower than the 199 minutes required for training with full 1024 dimensional SSL representations.


\section{Conclusion}

This paper proposes an AE-based optimization of SSL representations to reduce feature dimensions, improving the efficiency of dysarthric speech recognition in Kaldi's DNN-HMM ASR framework. By reducing input feature dimensions, we decrease ASR training and decoding time, optimize computational efficiency, and enable real-time applications. This approach lowers memory and processing requirements, making deployment feasible on resource-constrained devices while maintaining or improving accuracy. It also enhances model generalization, leading to more robust ASR performance in challenging speech conditions. Despite the obtained 8x reduction in training time, the AE-optimized HuBERT features did not seem to deteriorate the ASR performance.

\vspace{-0.2cm}
\bibliographystyle{IEEEbib}
\bibliography{strings,refs}

\end{document}